\documentclass[prl,floatfix,showpacs,amsmath,twocolumn]{revtex4}
\usepackage{amssymb}
\usepackage{graphics}

\begin{document}

\newcommand{\be}{\begin{eqnarray}}
\newcommand{\ee}{\end{eqnarray}}
\newcommand{\bea}{\begin{eqnarray}}
\newcommand{\eea}{\end{eqnarray}}
\newcommand{\bma}{\begin{subequations}}
\newcommand{\ema}{\end{subequations}}
\def\lR{l^2_{\mathbb{R}}}
\def\RR{\mathbb{R}}
\def\E{\mathbf e}
\def\D{\boldsymbol \delta}
\def\S{{\cal S}}
\def\T{{\cal T}}
\def\dd{\delta}

\title{Entanglement of formation for symmetric Gaussian states}

\author{G. Giedke$^{1,3}$, M.M. Wolf$^{1,2}$, O. Kr\"uger$^{2}$,
R.F. Werner$^{2}$, and J. I. Cirac$^{1}$}
\affiliation{\footnotesize (1) Max-Planck-Institut f\"ur
Quantenoptik,
Hans-Kopfermann-Str. 1, Garching, D-85748, Germany.\\
(2) Institut f{\"u}r Mathematische Physik, Mendelsohnstr.~3,
D-38106
Braunschweig, Germany\\
(3) Institut f{\"u}r Quantenelektronik, ETH Z{\"u}rich,
Wolfgang-Pauli-Stra{\ss}e, CH-8093 Z{\"u}rich, Switzerland}

\date{\today}

\begin{abstract}
We show that for a fixed amount of entanglement, two--mode
squeezed states are those that maximize
Einstein--Podolsky--Rosen--like correlations. We use this fact to
determine the entanglement of formation for all symmetric Gaussian
states corresponding to two modes. This is the first instance in
which this measure has been determined for genuine continuous
variable systems.
\end{abstract}

\maketitle

One of the main tasks of Quantum Information Theory is to quantify
the entanglement and the quantum correlations that quantum states
possess. For that, several entanglement measures have been
introduced during the last years \cite{MHorodecki}. In particular,
two of such measures stand out for their well defined physical
meaning: the entanglement of distillation and of formation (and
the corresponding asymptotic generalization, the entanglement
cost) \cite{Bennettlongpaper}. They quantify the entanglement of a
state in terms of the pure state entanglement that can be
distilled out of it \cite{PRHorodecki} and the one that is needed
to prepare it \cite{Wootters1}, respectively.

Despite a considerable effort, for the moment we can only evaluate
the entanglement of formation (EoF), or the entanglement of
distillation for a few sets of mixed states. The reason is that
these quantities are defined \cite{Bennettlongpaper} in terms of
an optimization problem which is extremely difficult to handle
analytically. Despite this fact, in a remarkable work Wootters
\cite{Wootters} managed to derive an analytical expression for the
EoF for all two--qubit states. The EoF has been also determined
for highly symmetric states (isotropic states \cite{Terhal}, and
Werner states \cite{symmetry}). These expressions are important
theoretical tools. From a more practical point of view, they can
be applied to quantify the entanglement created in current
experiments as well as to compare the capability of different
experimental set--ups. For low dimensional systems without
symmetries, one can still use numerical methods to determine the
EoF \cite{NumEoF}, although they are often not very efficient. For
infinite--dimensional systems, however, a numerical approach is
not feasible.

Among all quantum states in infinite dimensional systems, Gaussian
states play an important role in quantum information. From the
experimental point of view, they can be created relatively easily
\cite{squeezing}, and one can use them for quantum cryptography
\cite{Grangier} and quantum teleportation \cite{Kimble,Bachor}. On
the theoretical side, separability \cite{Geza} and distillability
\cite{Geza2} criteria for bipartite systems have been fully
developed. Moreover, pure Gaussian states are intimately related
to Heisenberg's uncertainty relation since they minimize such a
relation for position and momentum operators.

In this work we determine the EoF of all symmetric Gaussian states
of two modes. Those states arise naturally in several experimental
contexts. For example, when the two output beams of a parametric
down converter are sent through optical fibers \cite{squeezing},
or in atomic ensembles interacting with light \cite{expatoms}. In
order to determine the EoF, we connect  the entanglement of pure
states, as measured by the von Neumann entropy of the restriction,
with the kind of correlations established by Einstein, Podolsky
and Rosen (EPR) in their seminal paper \cite{EPR}. In fact, we
show that two--mode squeezed states \cite{2modesq} play a very
special role in this relation, since they are the least entangled
states for a given correlation of this type. This provides a new
characterization of two--mode squeezed states. Finally, we show
that the decomposition that leads to the EoF is a decomposition in
terms of Gaussian states.

We consider two modes, A and B, with corresponding Hilbert space
${\cal H}={\cal H}_A\otimes{\cal H}_B$ and canonical operators
$X_{A,B}$ and $P_{A,B}$. The two--mode squeezed states have the
form
 \be
 \label{Squeezed}
 |\Psi_s(r)\rangle := \frac{1}{\cosh(r)}\sum_{N=0}^\infty
 \tanh^N(r) |N\rangle_A\otimes|N\rangle_B,
 \ee
where $r> 0$ is the squeezing parameter, and $|N\rangle$ denotes
the $N$-th Fock state, i.e. $a^\dagger a|N\rangle_A =
N|N\rangle_A$, $b^\dagger b|N\rangle_B = N|N\rangle_B$, where
$a=(X_A+iP_A)/\sqrt{2}$ and $b=(X_B+iP_B)/\sqrt{2}$ are
annihilation operators.

In the following, we will denote by $\psi$ an arbitrary normalized
state in ${\cal H}$. We define its EPR--uncertainty as follows:
 \be
 \label{Delta}
 \Delta(\psi) := {\rm min}
 \left[1,\frac{1}{2}[\Delta^2_\psi(X_A-X_B)+\Delta^2_\psi(P_A+P_B)]\right],
 \ee
where, as usual, $\Delta^2_\psi(X):=\langle X\psi|X\psi\rangle-
\langle\psi|X|\psi\rangle^2$, setting $\Delta^2_\psi(X)=\infty$ if
$\psi$ is not in the domain of $X$. Clearly, $\Delta(\psi)\in
(0,1]$. This quantity measures the degree of non--local
correlations, and would be zero for the idealized state considered
by Einstein, Podolsky, and Rosen \cite{EPR}. For any state,
$\Delta(\psi)<1$ implies the existence of such non--local
correlations. Note that this condition is met only if at least one
of the uncertainties of $(X_A-X_B)/\sqrt{2}$ or
$(P_A+P_B)/\sqrt{2}$ lies below 1 (the standard quantum limit).
This implies that the corresponding states must possess a certain
squeezing. In fact, the two--mode squeezed states (\ref{Squeezed})
are standard examples of states displaying these correlations
since
 \be
 \label{DeltaPsi}
 \Delta[\Psi_s(r)] = e^{-2r}<1.
 \ee
Any value of $\Delta\in(0,1)$ is achieved by the two--mode
squeezed state with squeezing parameter
 \be\label{rDelta}
   r_\Delta := -\frac{1}{2}\ln(\Delta)
 \ee

The EPR--uncertainty of a given state $\psi$ is certainly related
to its entanglement. For pure states this last property is
uniquely quantified by the entropy of entanglement, $E(\psi)$,
which can be determined as follows. Let us write the Schmidt
decomposition of $\psi$ as \cite{notedecomp}
 \be
 \label{Psi}
 |\psi\rangle = \sum_{N=0}^\infty
 c_N |u_N\rangle_A\otimes |v_N\rangle_B,
 \ee
$\{u_N\}$ and $\{v_N\}$ are orthonormal bases in ${\cal H}_{A,B}$,
respectively and $c=(c_0,c_1,\ldots)\in {\cal C}$ where
 \[
 {\cal C}:= \{\,c \in \lR \mid ||c||=1, c_N\ge c_{N+1} \ge 0
 \,\forall N\,\}.
 \]
Then \cite{notelog}
 \be
 \label{EntPsi}
 E(\psi)=\E (c):=-\sum_{N=0}^\infty c_N^2 \log(c_N^2).
 \ee
Note that this quantity can be infinite for some states. For the
two--mode squeezed states (\ref{Squeezed}) we have
 \[
 E[\Psi_s(r)] = \cosh^2(r)
 \log[\cosh^2(r)]-\sinh^2(r)\log[\sinh^2(r)].
 \]

With the above definitions we can state the special
r{\^o}le that
two--mode squeezed states play in relation with EPR--correlations
and entanglement:

{\bf Proposition 1:} For all $\psi\in {\cal H}$, $E(\psi)\ge
E[\Psi_s(r_{\Delta(\psi)})]$.

In order to give a clear interpretation of this result, we
reformulate it in two equivalent forms: (i) For any given
$\Delta\in (0,1)$, $E[\Psi_s(r_\Delta)]=\inf_\psi \{E(\psi)\}$
with $\psi$ fulfilling $\Delta(\psi)=\Delta$; (ii) For any given
$E\in (0,\infty)$, $\Delta[\Psi_s(r)]= \inf_\psi \{\Delta(\psi)\}$
with $\psi$ fulfilling $E(\psi)=E$ and $r$ such that
$E[\Psi_s(r)]=E$. Note that the equivalence of this last
formulation is ensured by the fact that $E[\Psi_s(r)]$ and
$\Delta[\Psi_s(r)]$ are monotonically increasing and decreasing
functions of $r$, respectively. The first statement characterizes
two--mode squeezed states as the cheapest (regarding entanglement)
to achieve a prescribed EPR--uncertainty. The second one
characterizes them as those with maximal EPR--correlations
(minimal $\Delta$) for any given value of the entanglement.

In order to prove Proposition 1 we introduce two lemmas and the
following definition. Given $c\in {\cal C}$ we define
 \be
 \D(c):= 1+2 \sum_{N=0}^\infty  (c_N^2-c_Nc_{N-1})N.
 \ee
We have $\D(c)\le 1$ and $\D(c)=\Delta(\psi)$ whenever
$|u_N\rangle=|v_N\rangle=|N\rangle$ [cf. (\ref{Psi})].

{\bf Lemma 1:} For all $\psi$ with Schmidt decomposition
(\ref{Psi}), $\Delta(\psi) \ge \D(c)$.

{\em Proof:} Since $\D(c)\le 1$ we can restrict ourselves to
$\psi$ with $\Delta(\psi)<1$. Without loss of generality we can
assume that $\langle \psi|a|\psi\rangle=\langle
\psi|b|\psi\rangle=0$. Otherwise we can always find $\psi'$
fulfilling this condition, with the same Schmidt coefficients as
$\psi$ and with $\Delta(\psi')=\Delta(\psi)$ \cite{notedispl}. We
have
 \bea
 \Delta(\psi)&=&1+ \sum_{N=0}^\infty c_N^2
 (\langle u_N|a^\dagger a|u_N\rangle +
 \langle v_N|b^\dagger b|v_N\rangle)\nonumber \\
 &&- \sum_{N,M=0}^\infty c_N c_M (\langle
 u_N|a|u_M\rangle \langle v_N|b|v_M\rangle +
 c.c.).\nonumber\\
 &\ge& \min[Z(u),Z(v)],\nonumber
 \eea
where
 \bea
 \label{Z}
 Z(u):=&& 1+ 2\sum_{N=0}^\infty c_N^2 \langle u_N|a^\dagger a|u_N\rangle
 \nonumber\\
 &&- 2\sum_{N,M=0}^\infty c_N c_M |\langle
 u_N|a^\dagger|u_M \rangle|^2.\nonumber
 \eea
Without loss of generality let us assume that
$\min[Z(u),Z(v)]=Z(u)=:Z$. We can rewrite it as $
 Z=\sum_{N=0}^\infty \sum_{M=N+1}^\infty (c_N-c_M)^2 X_{N,M}$,
where $X_{N,M}:=|\langle u_N|a^\dagger|u_M \rangle|^2 + |\langle
u_M|a^\dagger|u_N  \rangle|^2$. Now, since $c\in {\cal C}$ we can
write $(c_N-c_M)^2 \ge \sum_{R=N}^{M-1} (c_R-c_{R+1})^2$, for
$M\ge N+1$, so that
 \bea
 Z &\ge& \sum_{R=0}^\infty (c_R-c_{R+1})^2
 \sum_{N=0}^R\sum_{M=R+1}^\infty X_{N,M},\nonumber\\
 &=& \sum_{R=0}^\infty (c_R-c_{R+1})^2 (R+1+2Y_R),\nonumber
 \eea
where
 \[
 Y_R := \sum_{N=0}^R \left[\langle u_N|a^\dagger_Na_N|u_N\rangle -
 \sum_{0=M\ne N}^R |\langle u_N|a^\dagger|u_M\rangle|^2\right],
 \]
with $a_N:=a -\langle u_N|a|u_N\rangle$. Now, using that $u_N\perp
u_M$ for $N\ne M$ we have $\langle u_N|a^\dagger|u_M\rangle =
\langle u_N|a_N^\dagger|u_M\rangle$ which, together with $
\sum_{0=M\ne N}^R |\langle
 u_N|a_N^\dagger|u_M\rangle|^2 \le \langle u_N|a^\dagger_Na_N|u_N\rangle$,
yields that $Y_R\ge 0$ for all $R$ and therefore
 \[
 \Delta(\psi) \ge  Z \ge \sum_{R=0}^\infty (c_R-c_{R+1})^2
 (R+1)\ge \D(c).\quad\mbox{\fbox{} }
 \]

Lemma 1 indicates that for a given set of Schmidt coefficients
$c\in {\cal C}$ EPR--correlations are maximized if the Schmidt
vectors are chosen to be  Fock states in the right order, i.e.
$|u_N\rangle=|v_N\rangle=|N\rangle$. Next we will show that for
fixed $\Delta$, the choice of Schmidt coefficients minimizing the
entropy of entanglement is given by those of a two--mode squeezed
state. Since the entropy and the EPR entanglement are explicitly
known functionals $\E(c)$ and $\D(c)$ on the sequences $c\in {\cal
C}$, this is a classical constrained variational problem.


{\bf Lemma 2:} For $\Delta\in(0,1)$, and any sequence $c\in {\cal
C}$ with $\D(c)=\Delta$, we have $\E(c)\geq \E(c^\Delta)\equiv
E[\Psi_s(r_\Delta)]$, where $c^\Delta_N\propto \exp(-Nr_\Delta)$
is the unique geometric sequence in $\cal C$ with
$\D(c^\Delta)=\Delta$.

{\em Sketch of proof:} We apply the method of Lagrange multipliers
for constrained minima to the infinitely many variables
$c_0,c_1\ldots$, leaving aside the technicalities of making this
rigorous.  These involve restricting $c$ to finite dimensional
spaces, then letting the dimension of the space tend to infinity,
and controlling the attained minima in this limit.

With a choice of Langrange multipliers $\mu$ and $\lambda>0$,
designed to simplify the expressions to come, we are thus looking
for stationary values $c\in\cal C$ of the functional
 \[
 F(c,\lambda,\mu):=\E(c)+\frac{\lambda}{2\ln(2)} [\D(c)-\Delta] +
 \frac{(\mu+1)}{\ln(2)}(||c||-1).
 \]
We obtain
 \be
 \label{cN}
 2c_N [N\lambda+\mu-\ln(c_N^2)]
 =\lambda[Nc_{N-1}+(N+1)c_{N+1}].
 \ee
where we have defined $c_{-1}=1$. One can immediately see that
$c_N>0$ and thus we can divide (\ref{cN}) by $c_N$ and subtract
the same expression but for $N+1$. Defining
$x_N:=c_{N+1}/c_{N}=:e^{-2r_N}\in (0,1]$ for $N=0,1,\ldots$ and
writing $\lambda=2r/\sinh^2(r)$ for some $r>0$ we find
 \be
 \label{XN}
 x_{N+1}=x_{N} - A_N - B_N,
 \ee
where $N=0,1,\ldots$ and
 \bma
 \bea
 A_N &=& \frac{4}{N+2} \left[\sinh^2(r_N)
 -\frac{r_N}{r}\sinh^2(r)\right],\\
 B_N &=& \frac{N}{N+2}\left[ \frac{1}{x_{N}}-\frac{1}{x_{N-1}}
 \right].
 \eea
 \ema
If we fix $r>0$ and $x_0$, we have three possibilities: (i) $x_0<
e^{-2r}$. Then, by induction, $x_N$ is decreasing, and will reach
some $x_N<0$ for finite $N$, which is impossible; (ii)
$x_0>e^{-2r}$. Then $x_N$ is increasing, and the normalization
condition for $c$ cannot be fulfilled. Hence we must have the
third possibility (iii) $x_0=e^{-2r}$, which implies that
$x_N=e^{-2r}$ for all $N$. Hence $c_N$ is a geometric sequence
$\propto \exp(-2Nr)$.\fbox

With the help of Lemmas 1 and 2 we are now in the position of
proving Proposition 1.

{\em Proof of Proposition 1:} Given $\psi\in {\cal H}$, if
$\Delta(\Psi)=1$ then it is trivial. Otherwise, using
(\ref{EntPsi}) and Lemma 2 we have
 \be
 E(\psi)=\E(c)\ge E[\Psi_s(r_{\D(c)})]\ge
 E[\Psi_s(r_{\Delta(\psi)})],
 \ee
where for the last inequality we have used $r_{\Delta(\psi)}\le
r_{\D(c)}$ [which follows from Lemma 1 and (\ref{DeltaPsi})] and
the fact that $E[\Psi_s(r)]$ increases monotonically with $r$.
\fbox{}

In the following, we will apply Proposition 1 to determine the EoF
of symmetric Gaussian states of two modes. For a given density
operator, $\sigma$, we define its covariance matrix (CM) $\gamma$
as usual,
 \be
 \gamma_{ij}:={\rm
 tr}[(R_iR_j+R_jR_i)\rho]-2{\rm tr}(R_i\rho){\rm tr}(R_j\rho),
 \ee
where $\{R_i, i=1,..,4\}:= \{X_A,P_A,X_B,P_B\}$. Up to local
unitary operations, it can always be written in the standard form
\cite{Geza}
 \be
 \gamma=\left(\begin{matrix}
 n &0&k_x&0\\ 0&n&0&-k_p \\k_x&0&m&0\\0&-k_p&0&m\\
 \end{matrix}\right).
 \ee
We will concentrate here on symmetric states, i.e. those which are
invariant under exchange of subindices A and B and therefore
fulfilling $m=n$. Without loss of generality we can choose
$k_x\geq k_p\geq 0$. In this case, $\gamma$ is a CM iff
$n^2-k_x^2\ge 1$ and describes an entangled state iff
$1>(n-k_x)(n-k_p)$ \cite{Geza}. Next we apply local (unitary)
squeezing transformations to yield the state in a more appropriate
form without changing its entanglement properties. In the
Heisenberg picture, the transformation multiplies (divides)
$X_{A,B}$ ($P_{A,B}$) by $[(n-k_p)/(n-k_x)]^{1/4}$. A simple
calculation gives
 \be
 \label{defdelta}
 \Delta(\sigma)=\sqrt{(n-k_x)(n-k_p)}=:\delta
 \ee
where $\Delta(\sigma)$ is defined analogously as in (\ref{Delta}).

Our goal is to determine the EoF of $\sigma$. This is defined as
$E_F(\sigma):= \inf_D {\cal E}(D)$, where the infimum is taken
with respect to all sets of the form $D=\{p_k,\psi_k\}$ which give
rise to a decomposition of $\sigma$, i.e.,
 \be
 \label{decrho}
 \sigma=\sum_k p_k |\psi_k\rangle\langle\psi_k|,
 \ee
where the $\psi_k\in {\cal H}$ are normalized and $p_k\ge 0$. Note
that the sum can run over continuous indices. For the set $D$ we
define
 \be
 {\cal E}(D):=\sum_k p_k E(\psi_k).
 \ee
We  call the set $D$ a decomposition of $\sigma$. A particular
decomposition $D_0$ of $\sigma$ is defined through
 \[
 \sigma \propto \int_{\RR^4} d\xi \, W(\xi)|\Psi_s(r_\dd)\rangle\langle
 \Psi_s(r_\dd)|W(\xi)^\dagger
 e^{-\frac{1}{4}\xi^T(\gamma-\gamma_{\dd})^{-1}\xi},
 \]
where $W(\xi)=e^{i\xi^T R}$ is the Weyl displacement operator and
$\gamma_\dd\le \gamma$ is the CM of the two--mode squeezed state
(\ref{Squeezed}) with squeezing parameter $r_\dd$. Since $W(\chi)$
are local unitary operators, we have ${\cal
E}(D_0)=E[\Psi_s(r_\dd)]$.

We also introduce the auxiliary function $f:(0,1]\to [0,\infty)$
 \be
 f(\Delta)=c_+(\Delta)\log[c_+(\Delta)]-
 c_-(\Delta)\log[c_-(\Delta)],
 \ee
where $c_{\pm}(\Delta):= ( \Delta^{-1/2}\pm \Delta^{1/2})^2/4$.
One can readily show that $f$ is a convex and decreasing function
of $\Delta$ and that
 \be
 \label{EPsi2}
 E[\Psi_s(r_\Delta)]=f(\Delta).
 \ee

{\bf Proposition 2:} $E_F(\sigma)=f[\sqrt{(n-k_x)(n-k_p)}]$.

{\em Proof:} We just have to prove that for any decomposition $D$,
${\cal E}(D)\ge f(\dd)$, where $\delta$ is given in
(\ref{defdelta}), since the decomposition $D_0$ already achieves
this value, i.e. ${\cal E}(D_0)=E[\Psi_s(r_\dd)]=f(\delta)$ [c.f.
Eq.~(\ref{EPsi2})]. For any decomposition we have
 \[
 {\cal E}(D)\ge \sum_k p_k f[\Delta(\psi_k)]\ge
 f\left[\sum_k p_k \Delta(\psi_k)\right]\ge
 f(\dd).
 \]
The first inequality is a consequence of Proposition 1 and
(\ref{EPsi2}). The second is due to the convexity of $f$. Finally,
the last one is a consequence of the fact that $\dd\ge \sum p_k
\Delta(\psi_k)$ (which can be easily checked by using the
Cauchy--Schwarz inequality) together with the fact that $f$ is a
decreasing function of its argument. \fbox{}

In summary, we have determined the EoF of symmetric Gaussian
states by establishing a connection between EPR--like correlations
and the entanglement of a state. The result implies that the
measured quantities in some of the recent experiments dealing with
atoms \cite{expatoms} and photons \cite{Glockl} not only qualify
entanglement but also quantify it. We expect that the methods
introduced here will allow to determine the EoF and other
properties of more general Gaussian states. The optimal
decomposition $D_0$ that gives rise to the EoF is a mixture of
Gaussian pure states, which means that those states are the
cheapest ones in terms of entanglement to produce symmetric
Gaussian states. Thus, it is tempting to conjecture that this is
also true for all Gaussian states. Finally, the results presented
here provide a new characterization of two--mode squeezed states
as those, which achieve a maximal EPR--like correlation for a
given value of the present entanglement.

We thank Frank Verstraete for discussions. This work has been
supported by the EU IST program (QUPRODIS and RESQ), and the
\emph{Kompetenznetzwerk Quanteninformationsverarbeitung der
Bayerischen Staatsregierung}. GG acknowledges funding by the
\emph{Alexander-von-Humboldt--Stiftung}.


\begin{thebibliography}{99}
\bibitem{notedecomp}
Note that some of the standard properties of state transformations
in finite dimensional Hilbert spaces are not longer true in
infinite dimensions. For example, two states with the same Schmidt
coefficients may not be transformed into each other by local
unitary transformations.
\bibitem{notedispl}
We take $|\psi'\rangle=W_A\otimes W_B |\psi\rangle$, where
$W_{A,B}$ are displacement (Weyl) operators fulfilling
$W_A^\dagger a W_A = a-\langle \psi|a|\psi\rangle$ and
$W_B^\dagger b W_B = b-\langle \psi|b|\psi\rangle$.
\end{thebibliography}
\end{document}